\documentclass[aps,prb,twocolumn,superscriptaddress]{revtex4-1}
\usepackage{amsmath, amssymb}
\usepackage{graphicx}
\usepackage{hyperref}
\usepackage{color}
\newcommand{\diad}[1]{\overset{\leftrightarrow}{#1}}

\begin{document}
\title{Photonic density of states enhancement in finite graphene multilayers}
\author{Ashley M. DaSilva}
\affiliation{Department of Physics, The University of Texas at Austin, Austin, Texas 78712-1192, USA}
\author{You-Chia Chang}
\affiliation{Applied Physics Program, The University of Michigan, Ann Arbor, Michigan 48109, USA}
\affiliation{Center for Ultrafast Optical Science, The University of Michigan, Ann Arbor, Michigan 48109, USA}
\author{Ted Norris}
\affiliation{Center for Ultrafast Optical Science, The University of Michigan, Ann Arbor, Michigan 48109, USA}
\affiliation{Department of Electrical Engineering and Computer Science, The University of Michigan, Ann Arbor, Michigan 48109, USA}
\author{Allan H. MacDonald}
\affiliation{Department of Physics, The University of Texas at Austin, Austin, Texas 78712-1192, USA}
\begin{abstract}
We consider the optical properties of finite systems composed of a series of graphene sheets separated by 
thin dielectric layers.  Because these systems respond as conductors to electric fields in the 
plane of the graphene sheets and as insulators to perpendicular electric fields, they can be 
expected to have properties similar to those of hyperbolic metamaterials.
We show that under typical experimental conditions graphene/dielectric multilayers have enhanced Purcell factors, and enhanced  
photonic densities-of-states in both the THz and mid-IR frequency range.  These behaviors can be traced to the 
coupled plasmon modes of the multi-layer graphene system.  We show that these results can be obtained with just a few layers of graphene. 
\end{abstract}
\maketitle

\section{Introduction} 

Hyperbolic metamaterials (HMMs) are artificially structured materials which 
have hyperbolic light dispersion, leading to an enhanced photonic  density-of-states.\cite{smith_electromagnetic_2003,lindell_bw_2001,smolyaninov_metric_2010,jacob_broadband_2012,jacob_engineering_2010} 
One approach that is used to design a HMM is to 
consider superlattices with alternating metal and dielectric layers and 
sub-wavelength periods.  When described as a homogenous
material using an effective medium approximation, the dielectric constants
of such a superlattice are:\cite{fang_optical_2009,kidwai_effective-medium_2012}
\begin{align}
\epsilon_{\parallel}=&\rho\epsilon_{d}+(1-\rho)\epsilon_{m}\\
\epsilon_{\perp}=& \left(\frac{\rho}{\epsilon_{d}}+\frac{1-\rho}{\epsilon_{m}}\right)^{-1}
\end{align}
where $\parallel$ and $\perp$ refer to the 
directions  parallel and perpendicular to the interfaces,
$\rho$ is the dielectric to metal thickness ratio, and $\epsilon_{d(m)}$ is the dielectric function of the dielectric (metal) constituent. 
By proper choice of materials, thickness ratio, and frequency, 
one can engineer a material with a hyperbolic dispersion relation,
{\em i.e.} a system in which $\epsilon_{\parallel}$ and $\epsilon_{\perp}$ 
have opposite signs over a wide range of frequencies.  

HMMs are important in photonic engineering,\cite{kildishev_engineering_2008} 
especially in applications to sub-wavelength 
imaging\cite{govyadinov_metamaterial_2006,smolyaninov_magnifying_2007} and confinement.\cite{liu_far-field_2007} Near-field thermal properties may also be engineered using HMMs for applications such as energy harvesting and thermal management.\cite{Guo_broadband_2012} 
One can also use HMMs to control luminescence via the Purcell effect,\cite{jacob_broadband_2012,kim_improving_2012} which reflects the dependence of spontaneous emission on the surrounding density of photonic states. 
In free space, the density of photonic states is proportional to $\omega^{2}$. 
In the presence of an interface, the density of states can be
enhanced by evanescent modes close to the interface. The size of the enhancement can be a few orders of magnitude, with most of the states localized close to the interface, leading to 
a more rapid excited state decay 
and enhanced photoluminescence of nearby atoms; this is the origin of the Purcell enhancement factor.
HMMs can have particularly strongly enhanced photonic densities-of-states.  

There are two types of HMMs depending on which components of the dielectric tensor are negative. 
Type I HMMs have a metal-like perpendicular dielectric constant ($\epsilon_{\perp}<0$, $\epsilon_{\parallel}>0$) 
while type II HMMs have a metal-like parallel dielectric constant  ($\epsilon_{\parallel}<0$, $\epsilon_{\perp}>0$). 
As an alternative to the alternating layer strategy,\cite{krishnamoorthy_topological_2012,jacob_engineering_2010} 
HMMs can also be constructed by embedding 
metallic nano-wires in a dielectric medium.\cite{tumkur_control_2012,kanungo_experimental_2010} 

Graphene, a monolayer of graphite, has long-lived long-wavelength plasmons which are tunable via gate voltage.\cite{ju_graphene_2011,bao_graphene_2012,grigorenko_graphene_2012,yan_tunable_2012,bludov_primer_2013} 
The possibility of gate tuning is one attractive feature of using graphene for the metallic 
layers in a HMM. 
Indeed, infinite graphene/dielectric stacks have recently been predicted to have a large Purcell factor and a negative\cite{iorsh_hyperbolic_2012} $\epsilon_{\parallel}$ and arrays of graphene ribbons have been predicted to perform favorably as a hyperlens.\cite{andryieuski_graphene_2012} Recent calculations of Fresnel coefficients and power spectra in the THz frequency regime provide further evidence of graphene's suitability as a component of HMMs.~\cite{othman_graphene-based_2013} Here we report a study of finite stacks of graphene layers in both the THz and mid-IR frequency ranges, and show that even for a small number of layers graphene/deielctric stacks retain desirable HMM properties, in particular an enhanced photonic density of states.\cite{[{}][{, contributed talk at the APS March Meeting, http://meetings.aps.org/link/BAPS.2013.MAR.R5.12}] dasilva_graphene_2013}
We emphasize that only in systems composed of a small number of graphene sheets will it actually be possible to 
modify the carrier densities and hence the plasmon frequencies
of individual graphene layers via the electric field effect.\cite{novoselov_electric_2004} 
(Screening prevents a back gate from influencing layers far from the substrate.\cite{datta_surface_2009}) 

We have also found that while having a dielectric between the graphene layers is important in order to prevent interlayer tunneling, its
direct role in modulating optical properties in tuning HMM effects 
for electromagnetic radiation in the THz regime is minimal. 
We find that graphene HMMs boast a large photonic density of states enhancement
for a wide range of frequencies, and that the properties are robust to the dielectric spacer thickness, Fermi energy, and elastic mean free path. We focus on wavevector-resolved transmission coefficient and photonic density of states, which show the presence of modes within the metamaterial which are evanescent in free space. The enlarged photonic density of states leads to a Purcell enhancement that is greatly improved relative to metal/dielectric layered HMMs.

We note that we have assumed equal carrier densities in all of the graphene layers.  This is an unrealistic approximation for multilayer graphene samples utilizing the elctric field effect to control the carrier density.\cite{sun_spectroscopic_2010} However, is possible to achieve approximately equal carrier densities by using doping techniques,\cite{ohta_controlling_2006,chen_charged-impurity_2008} and in this case we expect our assumption to be valid.

Our paper is organized as follows. 
We first describe the electromagnetic Green's functions and transfer matrices 
we use to perform calculations. 
We then characterize graphene HMMs by evaluating the reflection coefficients, 
showing that the anticipated HMM features are already realized at 
quite small graphene layer numbers. 
We then calculate the transmission coefficient and wavevector-resolved photonic density of states for a $N=6$-layer graphene-based HMM to illustrate the dependence of various properties on controllable parameters. 
Finally, we evaluate the photonic density of states and Purcell coefficient for finite graphene-based HMMs in both the THz and the mid-IR regime.

\section{Theoretical Formulation} 

Maxwell's equations in a uniform medium with dielectric constant $\epsilon$ and permittivity $\mu=1$
are conveniently solved using electromagnetic Green's functions defined by the differential equation\cite{joulain_definition_2003,agarwal_quantum_1975}
\begin{equation}
\nabla\times\nabla\times\diad{G}_{0}^{EM}(\mathbf{r})-\frac{\epsilon\omega^{2}}{c^{2}}\diad{G}_{0}^{EM}(\mathbf{r})=\delta(\mathbf{r}).\label{eqn:defG}
\end{equation}
This function is called the Diadic Green's function to reflect the 
property that a source oriented in one direction can in general result in electric and magnetic fields in any direction. In free space, the electric and magnetic Green's functions are the same. 
The electric and magnetic fields are obtained by integrating the electromagnetic Green's functions over the sources: 
\begin{align}
\mathbf{E}(\mathbf{r})=&\frac{4\pi i\omega}{c^{2}}\int d\mathbf{r}'\:\diad{G}_{0}^{EM}(\mathbf{r}-\mathbf{r}')\mathbf{J}(\mathbf{r}')\label{eqn:EfromG}\\
\mathbf{H}(\mathbf{r})=&\int d\mathbf{r}'\:\diad{G}_{0}^{EM}(\mathbf{r}-\mathbf{r}')\mathbf{M}(\mathbf{r}')
\end{align}
where $\mathbf{J}$ includes all free currents and $\mathbf{M}(\mathbf{r})=(4\pi/c)\nabla\times\mathbf{J}(\mathbf{r})$ 
can be thought of as the magnetization they produce.

The electromagnetic local density of states (LDOS) can also be calculated from the Green's function,\cite{joulain_definition_2003,agarwal_quantum_1975,girard_near_2005}
\begin{equation}\label{eqn:ldosDefn}
\rho(\omega,z)=\frac{\omega}{\pi c^{2}}\mbox{Im }\mbox{tr }\diad{G}
\end{equation}
where Im denotes the imaginary part and tr denotes the trace.  
In our planar geometry $\rho$ depends only on the coordinate $z$ 
which measures position relative to the graphene/dielectric multilayer. 
In a non-uniform medium, the electric and magnetic Green's functions will be different, and the total density of states is the sum of the electric and 
magnetic components, $\rho(\omega,z)=\rho^{E}(\omega,z)+\rho^{H}(\omega,z)$. \cite{joulain_definition_2003}

The Green's functions defined in Eqn.~\eqref{eqn:defG} are those of a uniform medium, while the expression for the LDOS (Eqn.~\eqref{eqn:ldosDefn}) depends on the Green's function in the non-uniform medium. We will assume that the top surface of an HMM of total thickness $L$ is located at $z=0$, and the regions $z>0$ and $z<-L$ is free space ($\epsilon=1$). 
The presence of the HMM can then be accounted for by writing the total electromagnetic field at $z>0$ 
as the sum of the incident and reflected parts.  
We find that the electric Green's function for $z>0$ can then be written as  
\begin{widetext}
\begin{align}
\diad{G}^{E}(k,z,z';\omega)&=\frac{i}{2K}\left[(\hat{s}\hat{s}+\hat{p}_{-}\hat{p}_{ - })e^{-iK(z-z')}\theta(z'-z)+(\hat{s}\hat{s} + \hat{p}_{ +}\hat{p}_{ +})e^{iK(z-z')}\theta(z-z') + (r_{s}\hat{s}\hat{s}+r_{p}\hat{p}_{ +}\hat{p}_{-})e^{iK(z+z')}\right]
\end{align}
\end{widetext}
where $k$ is the two-dimensional in-plane wavevector, 
$K=\sqrt{\epsilon\omega^{2}/c^{2}-k^{2}}$ is the out-of-plane wavevector, and $r_{\alpha}$ for $\alpha=s,p$ are the reflection coefficients for the two polarizations of EM waves. 
Here and below we set $\epsilon$ to $1$ for a HMM embedded in a vacuum.  
The polarization vectors are 
\begin{align}
\hat{s}=& \frac{1}{k}(k_{y}\hat{x}-k_{x}\hat{y})\nonumber\\
\hat{p}_{\pm}=&\frac{1}{\sqrt{k^{2}+K^{2}}}(\mp K\hat{k}+k\hat{z})\nonumber
\end{align}
denoting $s$ (TE) and $p$ (TM) polarized light. The $\pm$ index on $\hat{p}$ 
distinguishes upward moving and downward moving waves. 
The magnetic Green's function can be obtained from the electric Green's function by replacing $r_{s}\leftrightarrow r_{p}$.\cite{joulain_definition_2003} 
We now see that the Green's function for $z >0$, and therefore the LDOS, is determined 
solely by the reflection coefficients. The LDOS is
\begin{equation}
\rho(\omega,z)=\frac{\omega}{\pi c^{2}}\mbox{Re}\left\{\int\frac{dk}{2\pi}\frac{k}{2K}\left[4+\frac{4k^{2}}{\epsilon\omega^{2}/c^{2}}\left(r_{s}+r_{p}\right)e^{2iKz}\right]\right\}
\end{equation}
where $K=\sqrt{\omega^{2}/c^{2}-k^{2}}$ and $\epsilon=1$ when the HMM is in vacuum. The reflection coefficients, $r_{s}$ and $r_{p}$, are functions of $k$ and are explicitly provided below. 

Below we also consider the wave-vector resolved LDOS, $\rho(q,\omega,z)$, which 
separates contributions to the LDOS from different wavevectors: 
\begin{equation}
\rho(k,\omega,z)=\frac{\omega}{\pi c^{2}}\mbox{Re}\left\{\frac{2}{K}\left[1+\frac{c^{2}k^{2}}{\epsilon\omega^{2}}\:(r_{s}+r_{p}) e^{2iKz}\right]\right\}
\end{equation}
The vacuum LDOS can be recovered by setting the reflection coefficients to zero: 
\begin{align}
&\rho_{0}(k,\omega,z)\equiv \rho_{0}(k,\omega)=\frac{2\omega}{\pi c^{2}}\frac{\theta(\omega/c-k)}{\sqrt{\epsilon\omega^{2}/c^{2}-k^{2}}}\\
&\rho_{0}(\omega,z)\equiv\rho_{0}(\omega)=\frac{\omega^{2}}{\pi^{2}c^{3}}
\end{align}
which are independent of $z$.  Note that in the absence of an interface, there is 
no contribution to the DOS from wavevectors $k>\omega/c$.

The reflection coefficients $r_{s}$ and $r_{p}$ for $s$- and $p$-polarized light, respectively, are defined as the ratio of the reflected to the incident electromagnetic field at the interface. 
They are determined entirely by the boundary conditions imposed by Maxwell's equations: 
\begin{equation}
\left(\begin{array}{c}
E_{2\uparrow}\\ E_{2\downarrow}\end{array}\right)=M_{12}^{s(p)}\left(\begin{array}{c} E_{1\uparrow}\\ E_{1\downarrow}\end{array}\right)
\end{equation}
where the indices 1 and 2 refer to the region above and below a graphene sheet, 
and $\uparrow$ and $\downarrow$ denote the upward moving and downward moving modes, respectively. 
The matrices connecting the fields are,
\begin{align}
M_{12}^{s}=&\frac{1}{2}\left(\begin{array}{cc}
1+\frac{Q_{1}}{Q_{2}}-\frac{4\pi\sigma\omega}{c^{2}Q_{2}} & 1-\frac{Q_{1}}{Q_{2}}-\frac{4\pi\sigma\omega}{c^{2}Q_{2}}\\
1-\frac{Q_{1}}{Q_{2}}+\frac{4\pi\sigma\omega}{c^{2}Q_{2}} & 1+\frac{Q_{1}}{Q_{2}}+\frac{4\pi\sigma\omega}{c^{2}Q_{2}}\end{array}\right)\\
M_{12}^{p}=&\frac{1}{2}\left(\begin{array}{cc}
  1+\frac{\epsilon_{2}Q_{1}}{\epsilon_{1}Q_{2}}+\frac{4\pi\sigma Q_{1}}{\omega\epsilon_{1}}  &  1-\frac{\epsilon_{2}Q_{1}}{\epsilon_{1}Q_{2}}-\frac{4\pi\sigma Q_{1}}{\omega\epsilon_{1}}\\
  1-\frac{\epsilon_{2}Q_{1}}{\epsilon_{1}Q_{2}}+\frac{4\pi\sigma Q_{1}}{\omega\epsilon_{1}}  &  1+\frac{\epsilon_{2}Q_{1}}{\epsilon_{1}Q_{2}}-\frac{4\pi\sigma Q_{1}}{\omega\epsilon_{1}}\end{array}\right).
\end{align}
The two-dimensional conductivity of a 
graphene is given by\cite{hwang_dielectric_2007,falkovsky_optical_2007,stauber_optical_2008,mak_measurement_2008,abedinpour_drude_2011,jablan_plasmonics_2009,koppens_graphene_2011}
\begin{equation}
\sigma(\omega)=\frac{2e^{2}}{h}\left[i\frac{\varepsilon_{F}}{\hbar\omega+i\hbar/\tau}+\frac{i}{4}\ln\left\lvert\frac{\hbar\omega-2\varepsilon_{F}}{\hbar\omega+2\varepsilon_{F}} \right\rvert+\frac{\pi}{4}\theta(\hbar\omega-2\varepsilon_{F})\right]\label{eqn:sigma}
\end{equation}
Here $v_{D}$ is the Dirac velocity of graphene, $\varepsilon_{F}=\hbar v_{D}\sqrt{\pi n}$ is the Fermi energy as a function of carrier density $n$, $\tau$ is the transport time (which depends on the mobility), and $\theta(x)$ is the step function which 
specifies the threshold for 
interband transitions at large $\omega$. 
This expression ignores the nonlocal response, and is appropriate for $k\rightarrow 0$. 
In practice, this expression is found to work well away from the onset of interband transitions, which occurs when $\hbar\omega\approx 2E_{F}$.\cite{koppens_graphene_2011} Inserting a transfer matrix for each graphene layer, and one propagation matrix
\begin{equation}
P_{i}=\left(\begin{array}{cc}
e^{-iQ_{i}d_{i}} & 0\\
0 & e^{iQ_{i}d_{i}}\end{array}\right)
\end{equation}
for each dielectric layer of thickness $d_{i}$, the total transfer matrix is a product of the component matrices,
\begin{equation}
M^{s(p)}=\prod_{j}P_{j}M_{j-1,j}^{s(p)}
\end{equation}
The reflection coefficient, $r_{s(p)}$, and transmission coefficient, $t_{s(p)}$, are obtained from these expressions by solving 
\begin{equation}
\left(\begin{array}{c}0\\ t_{s(p)}\end{array}\right)=M^{s(p)}\left(\begin{array}{c}r_{s(p)}\\ 1\end{array}\right)
\end{equation}

\section{Optical Properties} 
The Purcell enhancement factor is defined as the ratio of the total radiation rate of a unit dipole 
source to the radiation rate of the dipole in vacuum:\cite{kidwai_effective-medium_2012,kidwai_dipole_2011}
\begin{align}
b=&1+\frac{3}{2\omega/c}\mbox{Re}\left\{\int_{0}^{\infty}\frac{dk\, k}{K}\left[f_{\perp}^{2}\frac{k^{2}r_{p}}{\omega^{2}/c^{2}}+{}\right.\right.\nonumber \\
&\qquad\qquad \left.\left.{}+\frac{1}{2}f_{\parallel}^{2}\left(r_{s}-\frac{K^{2}}{\omega^{2}/c^{2}}\: r_{p}\right)\right]e^{2iKz}\right\}\label{eqn:purcellfactor}
\end{align}
where $z$ is the surface to dipole distance, and $f_{\parallel}$ and $f_{\perp}$ are the components of the dipole along the directions parallel and perpendicular to the HMM layers, respectively.

\begin{figure}
\includegraphics[width=.9\columnwidth]{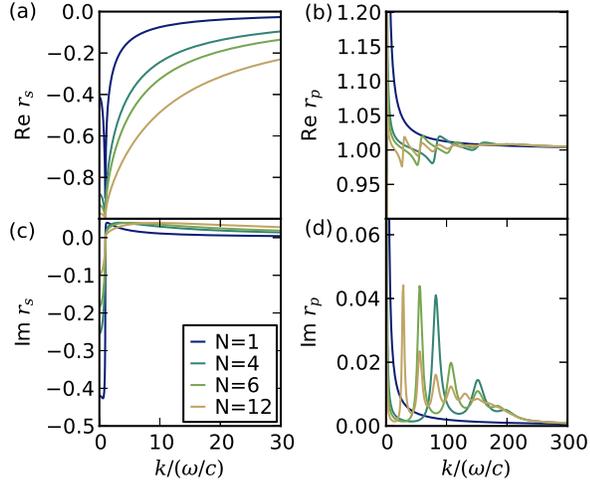}
\caption{Real (top panels) and imaginary (bottom panels) parts of the 
reflection amplitudes for $s$ (left panels) and $p$ (right panels) polarized EM waves on graphene HMMs
with different numbers of layers, $N$.  The parameters used for this calculation were frequency 
$f=1.0$ THz ($\hbar\omega=4.1$ meV), $n=4\times 10^{12}$ cm$^{-2}$ (Fermi energy $E_{F}=0.23$ eV)
in every layer, $\mu=50,000$ cm$^{2}$/Vs for graphene, $d=10$ nm for the dielectric layer 
thicknesses and $\epsilon=3.9$ for their dielectric constant.
}\label{rvsqvsN}
\end{figure}

In the effective medium approximation, the enhanced Purcell effect can be traced to the
nonzero imaginary part of the reflection coefficient in
the large $k$ limit, $k/(\omega/c)\rightarrow\infty$.\cite{kidwai_effective-medium_2012} 
In real systems, the finite period of the HMM limits the maximum value of $k$, but the signature of a 
nonzero imaginary part of the reflection coefficient for $k>\omega/c$ remains. 
These modes are evanescent in the vacuum on either side of the system, 
but propagate within the structure, as demonstrated by the enhanced transmission coefficient.\cite{cortes_quantum_2012}

\begin{figure}
\includegraphics[width=.9\columnwidth]{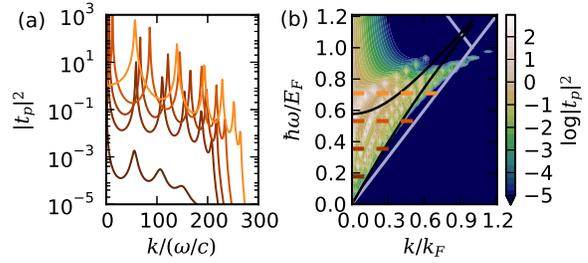}
\caption{(a) Magnitude squared of the transmission amplitudes
for $p$-polarized light on 6 layers of graphene HMM with the same material parameters as in Fig.~\ref{rvsqvsN} 
and $\omega$ ranging from $\hbar\omega=4.0$ meV ($f=1.0$ THz) to $\hbar\omega=165$ meV ($f=40$ THz), corresponding to the frequencies shown as colored lines in (b). 
(b) The logarithm of the magnitude squared of the transmission coefficient is shown as a function of frequency in units of Fermi energy and wavevector in units of Fermi wavevector. The Fermi energy is 233 meV. 
Overlayed on this plot are the threshold frequencies for inter band transitions (solid grey lines) 
and the frequency range of bulk plasmons in graphene superlattices (solid black lines).
}\label{Tvsk}
\end{figure}

In Fig~\ref{rvsqvsN} we plot the reflection coefficients for $s$- and $p$-polarized light as a function of wavevector normalized to frequency.
For $s$-polarized light, the number of graphene layers has little effect. For $p$-polarized light, a greater number of layers 
leads to more peaks of smaller magnitude while the general features remain intact, including the presence of 
a nonzero imaginary part of the reflection coefficient up to $k\sim 200\omega/c$. 
For the same parameters, figure~\ref{Tvsk} shows the transmittance {\em vs} parallel wave vector. 
Various values of frequency $\omega$ are considered. 
For both $s$- and $p$-polarized light, a significant fraction of the electromagnetic wave is transferred through the structure at $k<\omega/c$. 
This is due to the fact that the wavelength is much larger than the 
total thickness, $\lambda\gg Nd$, for $N=6$ layers each of thickness $d=10$ nm. 
We also observe finite transmission when $k>\omega/c$ for both polarizations.  As the frequency increases, the high-$k$ transmission coefficient decays more rapidly for $s$-polarized radiation compared to $p$-polarized radiation. This decay is due to the decreased wavelength, and is similar to the behavior of a uniform dielectric. 
On the other hand, the transmission coefficient for $p$-polarized radiation has several sharp peaks when $k>\omega/c$. A large value of $\lvert t_{p}\rvert^{2}$ corresponds to $t=E_{2\downarrow}/E_{1\downarrow}>1$. When $k>\omega/c$, the $\uparrow(\downarrow)$ labels correspond to the evanescent modes which decay to zero for $z\rightarrow -\infty$ and $z\rightarrow +\infty$, respectively. Any electromagnetic mode with $k>\omega/c$ must decay in the free space on either side of the HMM, so $E_{1\uparrow}\rightarrow 0$, yielding peaks at electromagnetic modes of the HMM. 

The peaks in transmission for $p$-polarized radiation are obtained in the same regime where there are peaks in the imaginary part of the 
reflection coefficient. (See Fig.~\ref{Tvsk}.)  Such features are not observed for $s$-polarized radiation which does not excite plasmon resonances since the electric field and parallel wavevector are perpendicular. 
These peaks reflect the 
coupled plasmon modes in the graphene layers,~\cite{shung_dielectric_1986} which have energies bounded  
by the dashed black lines in Fig~\ref{Tvsk}~(b).  These modes 
have been predicted and observed in 2DEG superlattices 
previously,~\cite{olego_plasma_1982,das_sarma_collective_1982,jain_plasmons_1985,pinczuk_discrete_1986,giuliani_charge-density_1983} and are usually discussed in terms of instantaneous intra and inter layer Coulomb interactions, an approximation
that is reliable in the large $k$ regime. 
In the context of HMMs these are sometimes called high-$k$ propagating 
modes~\cite{cortes_quantum_2012} and have application in subwavelength confinement
and imaging.~\cite{govyadinov_metamaterial_2006} 
As the frequency is increased, the transmission of $p$-polarized radiation is enhanced, with an 
optimal value at around $40$ THz (165 meV), close to the Fermi energy. 
For frequencies above the Fermi energy, the photon energy is above the maximum plasmon energy,~\cite{das_sarma_collective_1982,giuliani_charge-density_1983,jain_plasmons_1985,shung_dielectric_1986} and all high-$k$ modes disappear.

\begin{figure}
\includegraphics[width=.9\columnwidth]{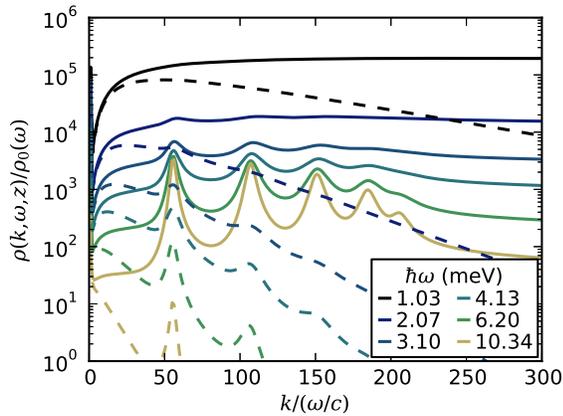}
\caption{The LDOS for a 6 layer graphene HMM with the same material parameters as Fig.~\ref{rvsqvsN} and different values of $\omega$ ranging from from $\hbar\omega=1.03$ meV ($f=0.25$ THz) to $\hbar\omega=10.34$ meV ($f=2.5$ THz), corresponding to the frequencies shown in the legend. The solid and dashed lines correspond to distances of $z=10$ nm and $z=1000$ nm above the surface of the HMM, respectively. Values of $(\omega/c)\ell_{mfp}\approx 0.003$ to $0.03$ for frequencies shown on this plot.
}\label{ldos_1}
\end{figure}

The desirable properties of HMMs stem from their enhanced LDOS at wavevector $k>\omega/c$. The LDOS will be a function of frequency $\omega$ and distance $z$ above the graphene HMM. The wavevector-resolved LDOS is shown in Fig.~\ref{ldos_1} for the same parameters as in Fig.~\ref{rvsqvsN} and Fig.~\ref{Tvsk} and for two different values of the distance $z$. 
As expected, the LDOS at large $z$ (dashed lines) decays more rapidly at large wavevector, due to the weak
influence of evanescent modes far from the surface of the HMM. 
We also notice that the LDOS enhancement is greater for smaller frequencies, in spite of 
the opposite trend in the transmission coefficient for $p$-polarization (See Fig.~\ref{Tvsk}~(b)). This is also an expected trend, and is due to the $1/\omega$ dependence of the LDOS which is apparent in the Purcell factor, Eqn.~\eqref{eqn:purcellfactor}. In addition to the overall larger LDOS, the peaks in the LDOS are smeared out for smaller frequencies because $\hbar \tau^{-1}\approx 0.7$ meV is held constant. 

One must be careful at large wavevectors ($k\approx k_{F}$) where the local approximation for the conductivity of graphene, Eqn.~\eqref{eqn:sigma}, 
becomes questionable.\cite{hwang_dielectric_2007}  
We have found that the LDOS decays before reaching $k=k_{F}$ for frequencies above $\approx 4$ meV at $10$ nm, and for all frequencies studied here at $1$ $\mu$m. At smaller frequencies and distances close to the interface the calculation may become unreliable unless nonlocal corrections to the conductivity are made. 
Nonlocal effects are known to provide a large wavevector cutoff of the LDOS enhancement.\cite{yan_hyperbolic_2012-1}
In the ballistic limit, $k\ell_{mfp}\gg 1$ ($\ell_{mfp}$ is the elastic mean free path for electrons), nonlocal effects are important when $\omega>v_{D}k$ where $v_{D}$ is the Dirac velocity of electrons. This limit reduces to $k/(\omega/c)<300$.

\begin{figure}
\includegraphics[width=.9\columnwidth]{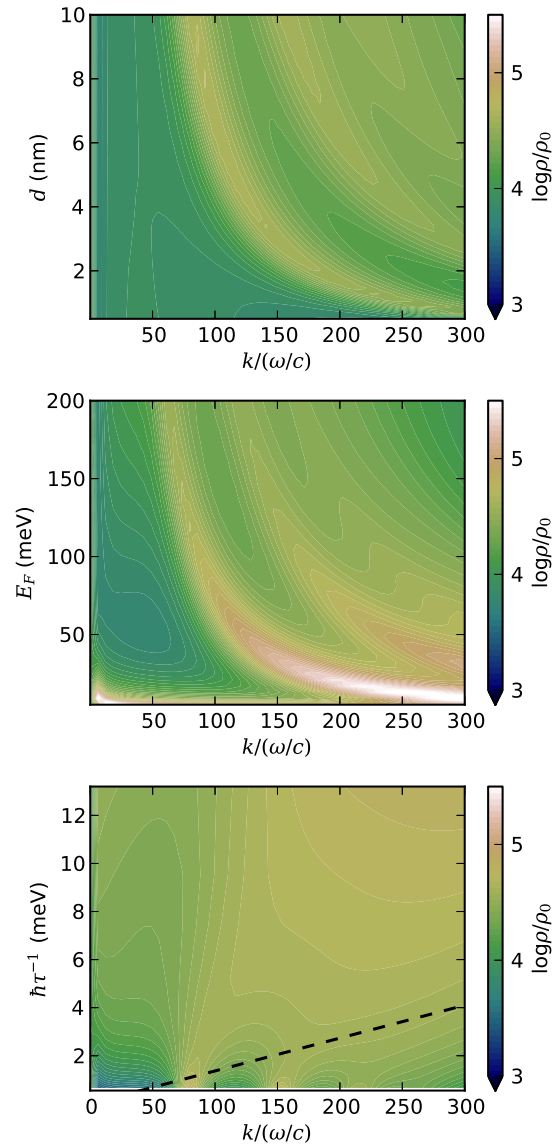}
\caption{The logarithm of the LDOS at $z=50$ nm as a function of parallel wavevector divided by vacuum wavevector, $k/(\omega/c)$, and (a) the distance between graphene sheets, $d$, for fixed Fermi energy $E_{F}=234$ meV and relaxation time $\tau=10^{-12}$ s; (b) Fermi energy, $E_{F}$, for fixed $d=10$ nm and $\tau=10^{-12}$ s; (c) inverse relaxation time, $\tau^{-1}$ for fixed $E_{F}=234$ eV and $d=10$ nm. For all plots, $\hbar\omega = 4.1$ meV and there are six layers of graphene separated by a dielectric with $\epsilon=3.9$. The dashed line in (c) corresponds to $k\ell_{mfp}=1$. In (a) and (b), $k\ell_{mfp}=3.7$ on the right hand side of the figure.
}\label{ldosContours}
\end{figure}

The other parameters in the model are the period of the graphene/dielectric superlattice, the carrier density, and the 
mobility of the graphene sheets. For simplicity, we have assumed that all graphene sheets have the same carrier density and mobility. 
The former assumption is unrealistic when carriers are induced by gates,
as mentioned previously, but does not influence properties in an essential way. 
Fig.~\ref{ldosContours}~(a) shows the dependence of the LDOS on both wavevector and period, $d$. The calculations were performed for a distance of $z=50$ nm from the surface. The LDOS is only weakly dependent on $d$ for features at small $k$; this is because for the THz regime under consideration, $d$ is always much smaller than $\lambda$ ($\lambda\sim 100$ $\mu$m for the frequencies studied here). However there is a noticeable enhancement of the high-$k$ features in the LDOS as $d$ decreases due to the dependence of the transfer matrix on the combination $kd$, which becomes comparable to 1 at $k/(\omega/c)\sim 500$ for $d=10$ nm and $k/(\omega/c)\sim 5000$ for $d=1$ nm. 
Fig.~\ref{ldosContours}~(b) shows the logarithm of the LDOS {\em vs.} Fermi energy and $k/(\omega/c)$. The dependence again is weak, but as the Fermi energy decreases, there is an enhancement when $\omega$ becomes comparable to $E_{F}$. 
For smaller Fermi energies, we expect that our local approximation for $\sigma(\omega)$ breaks down. 
In realistic multilayers that have carriers induced by gates, the density will normally be low in some layers.  
Fig.~\ref{ldosContours}~(c) shows the logarithm of the LDOS {\em vs} $\hbar\tau^{-1}$ and $k/(\omega/c)$. Again, the dependence is relatively weak, however there is a noticeable enhancement of the LDOS as $\tau$ decreases. This is probably due to stronger absorption in the graphene planes as the real part of the conductivity becomes larger.

\begin{figure}
\includegraphics[width=.9\columnwidth]{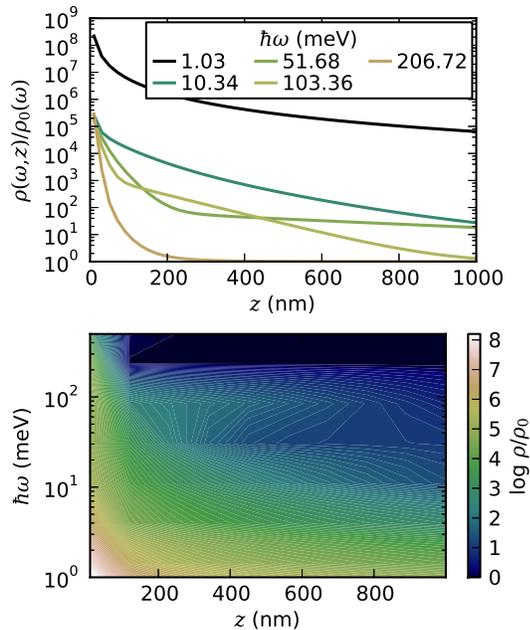}
\caption{(a) The integrated LDOS as a function of $z$ for several values of $\omega$ as indicated in the legend. (b) The logarithm of the integrated LDOS as a function of $\omega$ and $z$. For both (a) and (b), the system is six layers of graphene with $n=4\times 10^{12}$ cm$^{-2}$ and $\mu=50,000$ cm$^{2}$/Vs separated by $10$ nm of dielectric with $\epsilon=3.9$.
}\label{integratedLDOS}
\end{figure}

\begin{figure}
\includegraphics[width=.9\columnwidth]{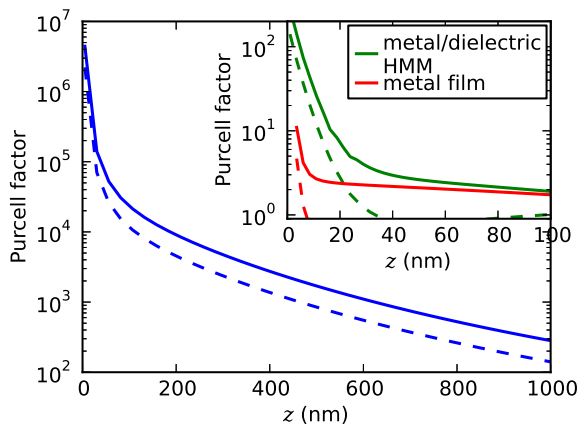}
\caption{The Purcell factor as a function of $z$ for a graphene HMM. The inset shows the Purcell factor as a function of $z$ for a metal/dielectric HMM and a metallic film. All three have 100 nm total thickness. The graphene HMM is composed of 10 layers of 10 nm each, having the same parameters as in Fig.~\ref{integratedLDOS} and at $\hbar\omega=4.1$ meV. The metal/dielectric HMM has 10 unit cells each composed of 5 nm of metal ($\epsilon_{m}=-27.5+0.31i$) and 5 nm of dielectric ($\epsilon_{d}=6.7$), and is at a frequency of $\hbar\omega=1.65$ eV. The metallic film has the parameters of the metallic part of the metal/dielectric HMM and is at the same frequency. In both the main figure and the inset, the solid lines are for a dipole oriented parallel to the HMM layers and the dashed lines are for a dipole perpendicular to the HMM layers.
}\label{purcellFactor}
\end{figure}

The Purcell factor depends not on the wavevector-resolved LDOS, but on the total LDOS, integrated over all wavevectors. In Fig.~\ref{integratedLDOS}(a) we show the integrated LDOS normalized to the vacuum LDOS {\em vs} $z$ for different values of $\omega$. As expected, the normalized LDOS decays away from the surface, and will reach a value of $1$ for $z\gg \lambda$. We also see that the normalized LDOS is larger for smaller values of $\omega$. Fig.~\ref{integratedLDOS}(b) shows the dependence of the normalized LDOS as a function of both $z$ and $\omega$ for six layers of graphene with $n=4\times 10^{12}$ cm$^{-2}$.  We again note that for small $\omega$ our local approximation for the conductivity  of graphene is questionable.

Next we calculate the Purcell factor a $10$ layer graphene HMM (total thickness $100$ nm) for the parameters $\hbar\omega=4.1$ meV, $n=4\times 10^{12}$ cm$^{-2}$, $\mu=50,000$ cm$^{2}$/Vs, and $\epsilon=3.9$. Fig.~\ref{purcellFactor} shows the Purcell factor calculated for the two orientations of a unit dipole: perpendicular to the surface (dashed) and parallel to the surface (solid lines). In the inset of Fig.~\ref{purcellFactor} we show the calculation of the Purcell factor for $100$ nm of both a metal/dielectric HMM (green) and metallic film (red).  The metal/dielectric HMM is composed of 10 unit cells of $5$ nm of metal with dielectric constant $\epsilon_{m}=-27.5+0.31i$, and $5$ nm of dielectric with dielectric constant $\epsilon_{d}=6.7$ at a wavelength of $750$ nm, parameters which are relevant for Ag/TiO$_{2}$ multilayers which have been used as typical HMMs for experiments and theory.\cite{cortes_quantum_2012,silva_optical_2004,johnson_optical_1972} The metallic film has the parameters for the metal at the same wavelength. Our results for the Purcell enhancement {\em vs} dipole distance are in agreement with those obtained in Ref.~\onlinecite{othman_graphene-based_2013} for a semi-infinite graphene-based HMM. We find that the Purcell factor for graphene is ehnanced at small distances compared to both the metal/dielectric HMMs and the metal films. However, one must note that the wavelength of the metal/dielectric HMM is necessarily different than the wavelength of the graphene HMMs. Metal/dielectric HMMs are limited to a frequency regime where the metal has a negative dielectric response (typically in the optical region of the electromagnetic spectrum), while the graphene HMMs are limited by the Fermi energy (typically in the THz to mid-IR region.) Note for example, at $z=1$ nm the metal/dielectric HMM has a Purcell factor $\sim 10^{2}$. Keeping the ratio of $z$ to free space wavelength $\lambda=2\pi c/\omega$ for the two metamaterials the same, the corresponding value of $z$ for the graphene-based HMM is $z\approx 400$ nm. We observe that the Purcell factor is $\sim 10^{3}$, a factor of 10 enhancement compared to the metal/dielectric HMM. 

\begin{figure}
\includegraphics[width=.9\columnwidth]{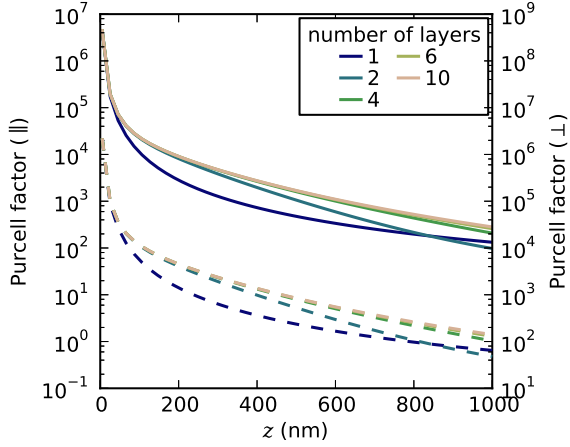}
\caption{The Purcell factor as a function of $z$ for a graphene HMM of 1, 2, 4, 6, and 10 layers, as denoted in the legend. The Purcell factors for parallel and perpendicular dipoles have been plotted against different axes for clarity. Solid lines correspond to a dipole oriented parallel to the surface, with Purcell enhancement given on the left axis, while dashed lines correspond to a dipole oriented perpendicular to the surface, with Purcell enhancement given on the right axis. The parameters of the graphene and dielectric are the same as in Fig.~\ref{integratedLDOS} and at $\hbar\omega=4.1$ meV. 
}\label{purcellFactorN}
\end{figure}

One obtains a large Purcell enhancement at small distances for small numbers of graphene layers. Fig.~\ref{purcellFactorN} shows a comparison of the Purcell factor for $N=1$, 2, 4, 6, and 10 layers of graphene. For one or two graphene layers the Purcell factor decays more rapidly. By the time there are $4$ graphene layers, the curves are almost identical up to $z=1000$ nm. This shows that graphene-based hyperbolic metamaterials are possible for few layers of graphene, as low as $N=4$. 

\begin{figure}
\includegraphics[width=.8\columnwidth]{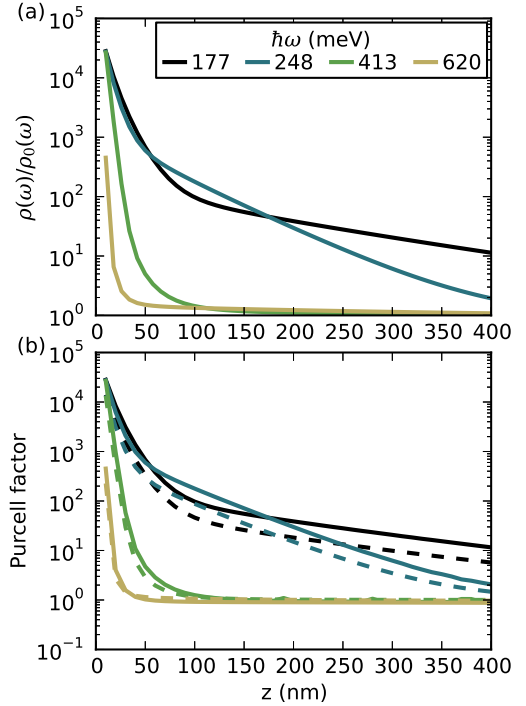}
\caption{Properties of a highly doped, low mobility graphene HMM in the mid-IR regime. In both panels,the graphene layers are separated by 10 nm of dielectric with $\epsilon=3.9$. The graphene layers have a carrier density of $5\times 10^{13}$ cm$^{-2}$ and a mobility of $\mu=1000$ cm$^{2}$/Vs. (a)  The integrated local density of states as a function of position $z$ above the top graphene layer. The wavelengths range from $3$ to $8$ $\mu$m, with energies $\hbar\omega$ indicated in the legend. (b) The Purcell factor of a unit dipole oriented parallel (solid) and perpendicular (dashed) to the surface of a 10-layer graphene metamaterial. The frequency of the oscillating dipole corresponds to those noted in the legend of panel (a).}\label{fig:CVD}
\end{figure}

So far our calculations are for high mobility graphene layers typical of exfoliated samples. Graphene grown by chemical vapor deposition (CVD) tends to have a lower mobility around 1000 cm$^{2}$/Vs. For such samples, it is beneficial to operate in the mid-infrared (mid-IR) regime, where 
the conductivity given by Eqn.~\ref{eqn:sigma} will still have a significant imaginary part despite the smaller relaxation time. Since the frequency regime for the HMM is limited by the Fermi level, it is necessary for such samples to be highly doped, to $10^{13}$ cm$^{-2}$ or larger carrier density. We have calculated the integrated LDOS and Purcell factor for highly-doped low mobility graphene at mid-IR wavelengths. Fig.~\ref{fig:CVD} shows that large LDOS and Purcell factor enhancements are predicted for these parameters. The Purcell factor is $~2$ orders of magnitude less than for the high mobility graphene in the THz regime; this is partially attributable to the ratio of $z/\lambda$ which is larger for mid-IR wavelengths. The Purcell factor remains, however, improved over that of the metal/dielectric HMM. 
We observe a crossover between wavelengths of $3$ and $5$ $\mu$m (177 and 248 meV) indicated by the much slower decay of the LDOS and Purcell factor with $z$. We attribute this cross-over to the transition from elliptical to hyperbolic isofrequency contour.\cite{krishnamoorthy_topological_2012} A simple Bloch theory\cite{constantinou_bulk_1986} for an infinite graphene-based metamaterial\cite{iorsh_hyperbolic_2012} shows that the transition from an effective permittivity $\epsilon_{\parallel}>0$ to $\epsilon_{\parallel}<0$ should occur in exactly this regime, while $\epsilon_{\perp}$ remains positive.

We should stress here the non-negligible role of loss in the enhancement of the Purcell factor. A lossy material need not be hyperbolic in order to produce a large Purcell enhancement: for example the metallic film in the inset of Fig.~\ref{purcellFactor} has a larger Purcell enhancement than a metal/dielectric HMM at small enough distances. For hyperbolic systems, instead we see an enhanced Purcell coefficient over a longer distance.

\section{Summary}

In conclusion, we have found that thin graphene stacks are HMMs in the THz to mid-infrared regime for a wide range of parameters.  We have studied the high-$k$ propagating modes as well as the wavevector resolved local density of states for graphene stacks and find an enhancement of both quantities at wavevectors which are evanescent in vacuum. This implies that enhanced near-field effects including sub-wavelength imaging and confinement of light may be possible. We also calculate the Purcell factor for both our graphene HMMs, and HMMs composed of metal/dielectric stacks. We find that the graphene HMMs perform very well compared to this benchmark at both the THz and mid-IR wavelengths. The frequency range of the graphene HMM is limited by the Fermi energy, $\hbar\omega\lesssim\varepsilon_{F}$, and so the graphene must be highly doped for mid-IR applications. We observe a transition from high Purcell enhancement to low Purcell enhancement around 3-5 $\mu$m for the low mobility, highly doped graphene which we attribute to a transition from hyperbolic to elliptical isofrequency contour.

\section{Acknowledgments}
Work at the University of Texas was supported by the Welch foundation under grant TBF1473 and the DOE Division of Materials Sciences and Engineering under grant DE-FG03-02ER45958. Michigan researchers were supported by the NSF MRSEC program under DMR 1120923.

\bibliographystyle{apsrev4-1}
%

\end{document}